\documentclass[iicol,Numbered]{sn-jnl}

\usepackage{subfig}
\usepackage{graphicx}%
\usepackage{physics}
\usepackage{multirow}%
\usepackage{amsmath,amssymb,amsfonts}%
\usepackage{amsthm}%
\usepackage{mathrsfs}%
\usepackage[title]{appendix}%
\usepackage{xcolor}%
\usepackage{textcomp}%
\usepackage{manyfoot}%
\usepackage{booktabs}%
\usepackage{algorithm}%
\usepackage{algorithmicx}%
\usepackage{algpseudocode}%
\usepackage{listings}%
\usepackage{booktabs}
\usepackage{multirow}
\usepackage{array}
\usepackage{cite}

\usepackage{graphicx}
\usepackage{textcomp}
\definecolor{red}{rgb}{0, 0, 0}
\usepackage{amsmath}
\usepackage{algorithm,algcompatible}
\usepackage{caption}
\usepackage{}
\usepackage{amsthm}

\usepackage{enumitem}
\usepackage{float}

\begin{document}

\title{Cloud-based Semi-Quantum Money}
\author[1]{\fnm{Yichi} \sur{Zhang}}\email{lionel.zhang@connect.ust.hk}
\equalcont{These authors contributed equally to this work.}

\author[1]{\fnm{Siyuan} \sur{Jin}}\email{siyuan.jin@connect.ust.hk}
\equalcont{These authors contributed equally to this work.}

\author[1]{\fnm{Yuhan} \sur{Huang}}\email{yhuangfv@connect.ust.hk}

\author[1]{\fnm{Bei} \sur{Zeng}}\email{bzeng@ust.hk}

\author*[1]{\fnm{Qiming} \sur{Shao}}\email{eeqshao@ust.hk}

\affil*[1]{\orgname{Hong Kong University of Science and Technology}, \orgaddress{\city{Hong Kong}}}

\abstract{In the 1970s, Wiesner introduced the concept of quantum money, where quantum states generated according to specific rules function as currency. These states circulate among users with quantum resources through quantum channels or face-to-face interactions. Quantum mechanics grants quantum money physical-level unforgeability but also makes minting, storing, and circulating it significantly challenging. Currently, quantum computers capable of minting and preserving quantum money have not yet emerged, and existing quantum channels are not stable enough to support the efficient transmission of quantum states for quantum money, limiting its practicality. Semi-quantum money schemes support fully classical transactions and complete classical banks, reducing dependence on quantum resources and enhancing feasibility. To further minimize the system's reliance on quantum resources, we propose a cloud-based semi-quantum money (CSQM) scheme. This scheme relies only on semi-honest third-party quantum clouds, while the rest of the system remains entirely classical. We also discuss estimating the computational power required by the quantum cloud for the scheme and conduct a security analysis.}


\maketitle
\section{Introduction}

\subsection{Research Background}
In recent years, advancements in quantum computing and information theory have renewed interest in quantum money. Researchers have developed new models, such as public-key quantum money, which leverage the complexity of quantum problems to ensure security. Progress in qubit technologies, including trapped ions, superconducting circuits, and photonic systems, demonstrates the feasibility of creating and controlling quantum states with high precision.

However, practical quantum money faces several challenges. Quantum states are fragile and susceptible to decoherence, often leading to information loss. Scalability is another major issue, as current quantum systems are small and prone to errors.

The concept of semi-quantum money introduces a new type of quantum money that keeps classical verifiability but changes the minting process. Unlike traditional quantum money, where the bank mints the money using quantum algorithms and sends it via quantum channels, semi-quantum money involves a protocol where both the bank and the user collaborate using classical resources.

In semi-quantum money, the user generates the quantum money state using instructions provided by the bank. This process is designed to ensure that the user cannot replicate the state to create additional money. The verification of semi-quantum money is conducted using classical interactive protocols between the user and the bank, avoiding the need for quantum communication or computation during verification.

\begin{table*}[ht]
\centering
\footnotesize
\caption{Comparison of Quantum Money Protocol \label{table: quantum_money}}
\begin{tabular}{@{}cccccc@{}}
\toprule
\textbf{Protocol} & \textbf{\begin{tabular}[c]{@{}c@{}}Bank\\ Node\end{tabular}}  & \textbf{\begin{tabular}[c]{@{}c@{}}Client\\ Node\end{tabular}} & \textbf{\begin{tabular}[c]{@{}c@{}}Bank-Client\\ Channel\end{tabular}} & \textbf{\begin{tabular}[c]{@{}c@{}}Client-Client\\ Channel\end{tabular}} & \textbf{Cloud} \\ \midrule
\cite{wiesner1983conjugate} & \color{red}{Quantum}     & \color{red}{Quantum}      & \color{red}{Quantum}                                                                    & \color{red}{Quantum}                                                                      & No      \\
\cite{aaronson2009quantum} & \color{red}{Quantum}     & \color{red}{Quantum}      & \color{red}{Quantum}                                                                    & \color{red}{Quantum}                                                                      & No      \\
   \cite{shmueli2022public} & Classical     & \color{red}{Quantum}      & Classical                                                                    & \color{red}{Quantum}                                                                      & No      \\
   \cite{shmueli2022semi} & Classical     & \color{red}{Quantum}      & Classical                                                                    & Classical                                                                      & No      \\
    Our Work & Classical     & Classical      & Classical                                                                    & Classical                                                                      & Yes      \\
   \bottomrule
\end{tabular}
\end{table*}

Table \ref{table: quantum_money} shows the comparison of different quantum money protocols. Quantum money have been an actively researched topic for several decades. Public-key quantum money were first conceptualized by \cite{wiesner1983conjugate} in the 1960s, where there is a bank that generates a secret-basis quantum state allowing easy verification of the issued quantum money. The quantum money is unique and can not be replicated without access to secret serial numbers. However, the system requires that users have quantum computers and quantum channels with the banks to verify each transaction.

Many researchers have subsequently suggested different quantum money schemes, aiming to facilitate verification by any entity possessing a quantum device \cite{aaronson2009quantum, aaronson2012quantum, zhandry2021quantum}. Unfortunately, the methods proposed by \cite{aaronson2009quantum} and \cite{aaronson2012quantum} were completely broken \cite{lutomirski2009breaking, pena2015algebraic}. For example, \cite{zhandry2021quantum} sought to resolve these issues by proposing a modification, which incorporated an indistinguishability obfuscator into the protocol. Despite this, \cite{roberts2021security} highlighted that the protocol's hardness assumption was flawed. 

In a subsequent response, \cite{shmueli2022public} proposed that an ideal quantum token should rely on local quantum computation and classical communication, obviating the need for quantum communication. To this end, they integrated a homomorphic encryption technique into the quantum money mining process, thereby facilitating secure and feasible local mining.

Our approach provides a significant departure from purely quantum money schemes, as it leverages classical resources for both minting and verification while maintaining security properties comparable to quantum systems.

\subsection{Motivation}
At the current technological level, the cost of manufacturing and maintaining quantum computers is extremely high, making it impractical for ordinary people to own and use them directly. Looking ahead, as quantum computers gradually enter practical use, cloud quantum computing services provided by a few large institutions are likely to become mainstream. This trend is similar to the evolution of classical computers: from early large centralized computers to later personal computers, which took several decades. By analogy, using cloud quantum computers to support the use of semi-quantum money could greatly accelerate its practical adoption.

However, the high dependence of semi-quantum money on security poses new challenges. Traditional security mechanisms may appear fragile in the face of quantum computing. Therefore, ensuring that cloud-based semi-quantum money is not compromised by unreliable third parties becomes an urgent issue. Specifically, we need to explore whether there is a solution that, under the premise of ensuring security, can utilize third-party quantum cloud services along with classical users, banks, and communication channels to implement a feasible cloud-based semi-quantum money system.

The core of this paper is to address this question. We will first review the basic concepts of semi-quantum money and the existing technical challenges, then analyze the historical experience of the evolution from centralized computing to personal computing, and explore the potential of cloud quantum computing in promoting the practical use of semi-quantum money. On this basis, we will design and verify a secure cloud-based semi-quantum money scheme. Through theoretical analysis and experimental verification, we aim to propose a feasible solution to lay the foundation for the widespread application of semi-quantum money.

\subsection{Results}
Our main result is the protocol of the cloud-based semi-quantum money (CSQM). We first formally propose the specific steps of this protocol and use a small 4-qubit demonstration to help understand it. Then, we conducte a rigorous analysis of the security of this protocol based on information-theoretic security and estimate the feasibility of this protocol based on the current hardware level.

\section{Preliminary}
This section introduces the cryptography primitives and notions borrowed from the original papers \cite{mahadev2020classical, shmueli2022semi, garg2016candidate}.
\subsection{Quantum Leveled Fully Homomorphic Encryption}

\begin{itemize}
    \item $fhek \leftarrow QHE.KeyGen(1^{\lambda}, 1^L)$: A classical probabilistic polynomial-time (PPT) algorithm samples a classical secret key $fhek$ from given security parameter $\lambda \in \mathbb{N}$ and circuit depth $L \in \mathbb{N}$.
    \item $Z^zX^x\ket{\Psi}:= \ket{\Psi}^{x,z} \leftarrow QHE.QOTP_{z, x}(\ket{\Psi})$: A quantum polynomial-time algorithm given random string $x, z \in \{0, 1\}^{\lambda}$ and a $\lambda$-qubit quantum state $\ket{\Psi}$, outputs $Z^zX^x\ket{\Psi}:= \ket{\Psi}^{x,z}$.
    \item $x\oplus m \leftarrow QHE.OTP_x(m)$: A classical algorithm encrypts message $m \in \{0,1\}^*$ with $x$ that $|m| = |x|$.
    \item $ct_x \leftarrow QHE.Enc_{fhek}(x)$: A classical PPT algorithm encrypts bit(s) in classical tensor $x$ bit-wisely and outputs the classical ciphertext $ct_x$.
    \item $(C\ket{\Psi})^{x',z'}, ct_{x', z'} \leftarrow QHE.Eval((\ket{\Psi}^{x,z}, ct_{x,z}), C)$: A quantum polynomial-time algorithm with input a general quantum circuit $C$, a quantum one-time-pad encrypted state $\ket{\Psi}^{x,z}$ and the ciphertext $ct_{x,z}$. The algorithm will output a new quantum state $(C\ket{\Psi})^{x',z'}$ and a classical ciphertext $ct_{x',z'}$.
    \item $x \leftarrow QHE.Dec_{fhek}(ct_x)$: A PPT algorithm uses the secret key $fhek$ to decrypt the ciphertext $ct_{x}$ bit-wisely.
\end{itemize}

\subsection{Quantum Homomorphic Encryption Circuit}
We use quantum fully-homomorphic encryption (QFHE) \cite{mahadev2020classical} as the cryptography instantiation, which is based on the standard hardness assumption of learning with errors (LWE). The classical sender sends a one-time-padded quantum state (can be described classically) $\ket{\psi}^{(x,z)}$, the ciphertext $ct_{(x,z)}$ of the one-time-pad and a general quantum circuit $C$ to a receiver with a quantum computer. Then the receiver can output $(C\ket{\psi})^{(x',z')}$ along with the ciphertext $ct_{(x',z')}$. Since a general quantum circuit can be divided into layers of Pauli, Clifford and Toffoli gates. If the circuit $C$ contains only Pauli and Clifford gates, the transform from $ct_{x,z}$ to $ct_{x',z'}$ is determinate  \cite{childs2001secure}. And after applying Toffoli gate, the transformation of $ct_{x,z}$ is random. Since the process is non-trivial, we will introduce the QFHE Toffoli gate in the next section. Our protocol will make use of the randomness to mint the anti-counterfeit quantum state as semi-quantum money.

\subsubsection{Encrypted CNOT Operation} 
\begin{figure*}
    \centering
    \includegraphics[width=1\textwidth]{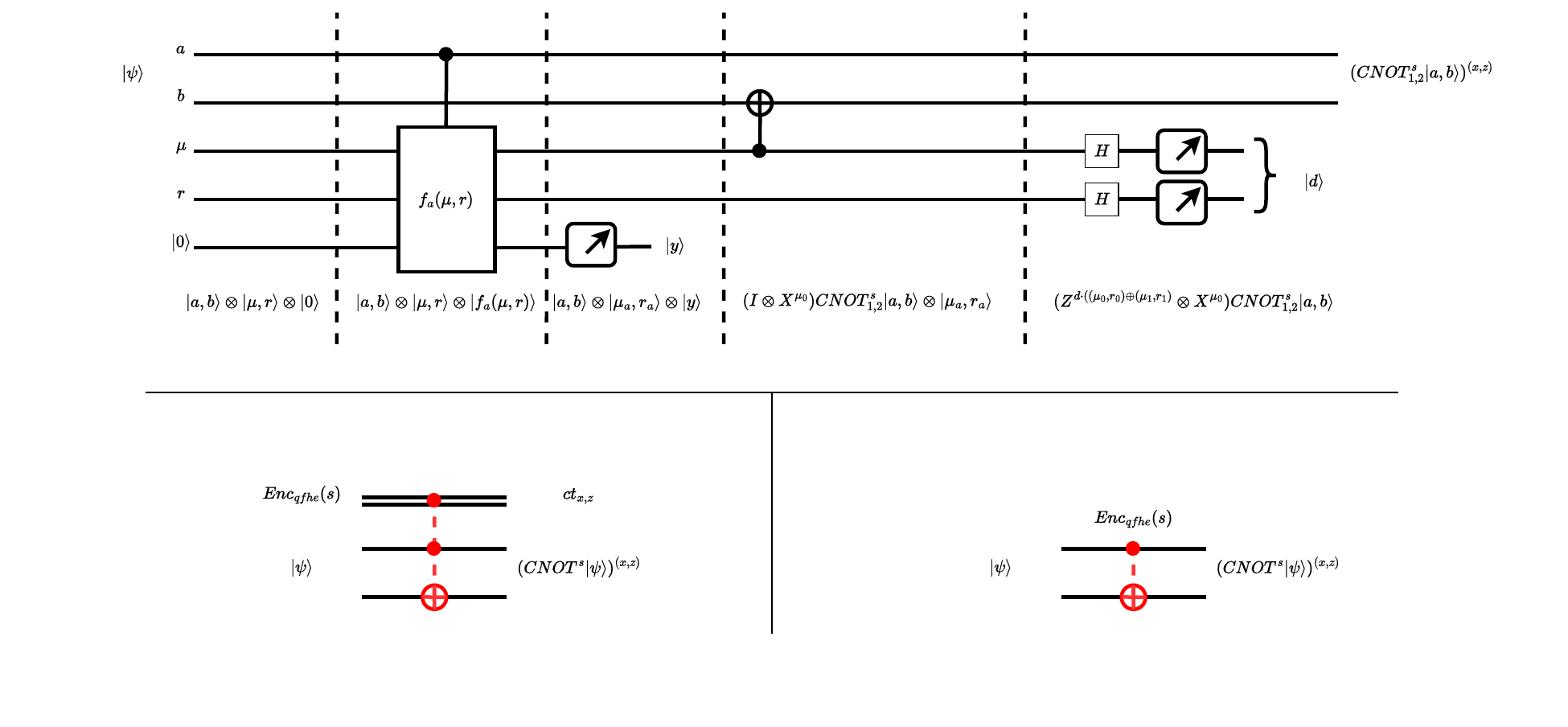}
    \caption{\textbf{Top}: the quantum circuit of encrypted CNOT gate defined in \cite{mahadev2020classical}. \textbf{Lower left}: the abstract form of the encrypted CNOT operation. \textbf{Lower right}: the simplest form of the encrypted CNOT operation, which will be used in other diagrams.}
    \label{CNOTs}
\end{figure*}

Encrypted CNOT operation is the key module of QFHE Toffoli operation. The sender samples a pair of injective functions $f_0(\mu,r), f_1(\mu,r)$ as the encryption of a bit $s$ and a 2-qubit quantum state $\ket{\psi}=\ket{a,b}$, then send them to the receiver. The pair of $f_0(\mu,r), f_1(\mu,r)$ has below properties:
\begin{enumerate}
    \item Injective
    \item Easy to be convert with a trapdoor $t_A$
    \item Computationally difficult to find any pair of preimages $(\mu_0,r_0), (\mu_1,r_1)$ with the same image $f_0(\mu_0,r_0) = f_1(\mu_1,r_1) =y$ without $t_A$
    \item For every $f_0(\mu_0,r_0) = f_1(\mu_1,r_1) =y$, $\mu_0 \oplus \mu_1 = s$
\end{enumerate}
Figure \ref{CNOTs} shows the quantum circuit of the encrypted CNOT operation and the abstract form that will be referred in further diagram. In the end of the encrypted CNOT operartion, the receiver will output the quantum state $(CNOT^s \ket{\psi})^{(x,z)}$ and $ct_{(x,z)}$, the ciphertext of the pad $(x,z)$.

\subsubsection{Toffoli Operation on Encrypted Quantum State}
\begin{figure*}
    \centering
    \includegraphics[width=1\textwidth]{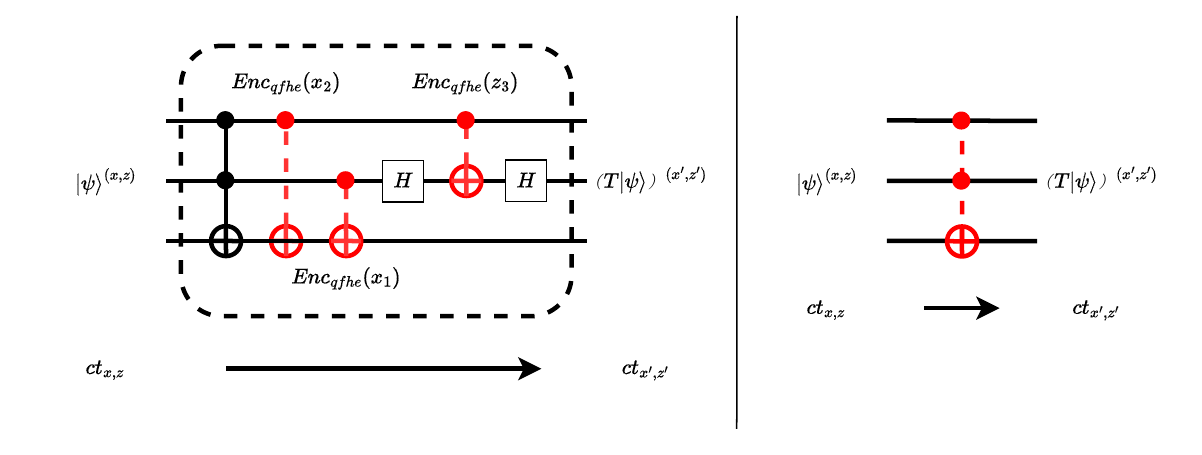}
    \caption{\textbf{Left}: the quantum circuit of the QFHE Tofolli operation. \textbf{Right}: the abstract form of QFHE Tofolli operation.}
    \label{toffoli}
\end{figure*}

When applying a Toffoli gate directly on a OTP quantum state $Z^z X^x \ket{\psi}$ we can get $TZ^zX^x\ket{\psi} = (TZ^zX^xT^\dagger) T\ket{\psi}$, where $TZ^zX^xT^\dagger = T Z^{z_1} X^{x_1} \otimes Z^{z_2} X^{x_2} \otimes Z^{z_3} X^{x_3} T^\dagger$.\\ And $T Z^{z_1} X^{x_1} \otimes Z^{z_2} X^{x_2} \otimes Z^{z_3} X^{x_3} T^\dagger = CNOT^{x_2}_{1,3} CNOT^{x_1}_{2,3}\hat{Z}^{z_3}_{1,2}Z^{z_1+x_2z_3}X^{x_1} \otimes Z^{z_2+x_1z_3} X^{x_2} \otimes Z^{z_3} X^{x_1x_2+x_3}$. Based on the expansion, we apply the encrypted CNOT gates (Figure \ref{toffoli}) to reveal the Pauli layer $T\ket{\psi}^{(x',z')}$ along with the new ciphertext $ct_{(x',z')}$.

\subsection{Indistinguishability Obfuscation}
We call a PPT algorithm $iO$ is an  \textit{indistinguishability obfuscator} over a circuit class $\{C_{\lambda}\}$ that satisfied: 
\begin{itemize}
    \item For all security parameters $\lambda \in \mathbb
{N}$, all $C \in C_{\lambda}$ and all input $x$:
$$Pr[O(x) = C(x): O\leftarrow iO(\lambda, C)] = 1$$
    \item For any PPT distinguisher $D$, all pairs $C_0, C_1 \in C_{\lambda}$, there exists a negligible function $negl$ that satisfied:
    \begin{align*}
        |Pr[D(iO(\lambda, C_0))=1] & - Pr[D(iO(\lambda, C_1))=1]| \\
        & \leq negl(\lambda)
    \end{align*}
\end{itemize}

\section{Main Result}
\subsection{Cloud-based Semi-quantum Money Protocol}
\begin{figure}[ht]
    \centering
    \subfloat[Mint\label{fig:mint}]{\includegraphics[width = 0.5 \textwidth]{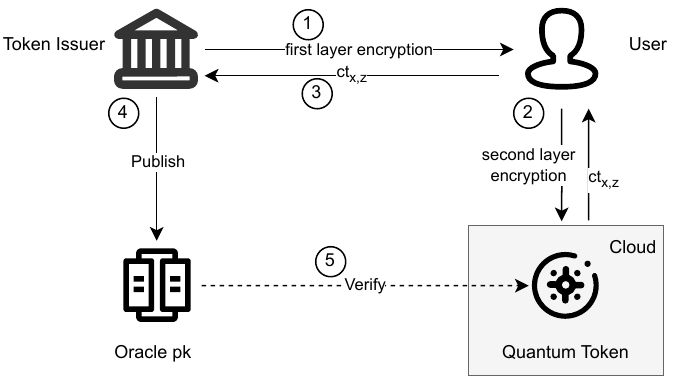}}\\
    \subfloat[Redeem\label{fig:redeem}]{\includegraphics[width = 0.5 \textwidth]{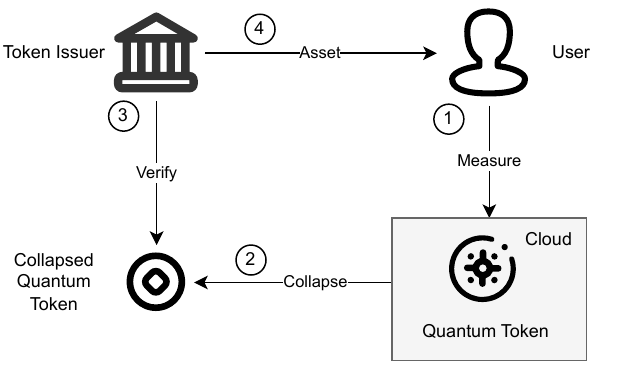}}\\
     \subfloat[Circulation\label{fig:circulation}]{\includegraphics[width = 0.475 \textwidth]{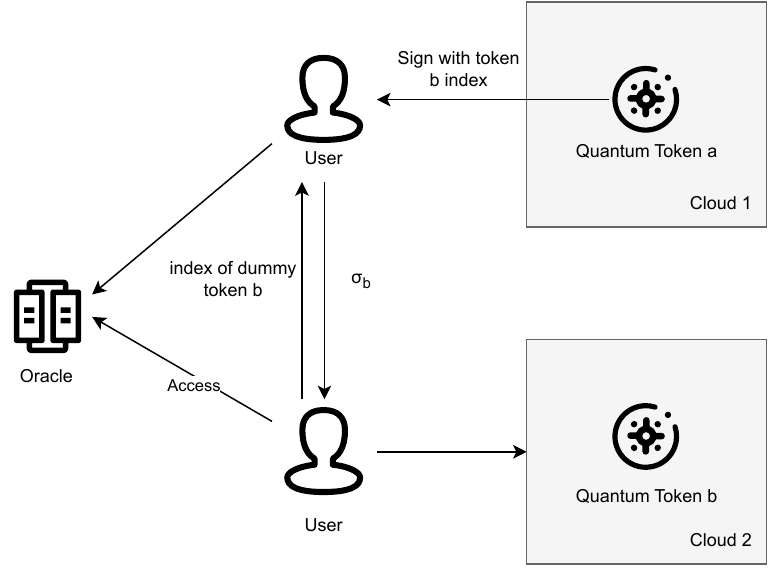}}
    \caption{Quantum Token Process.}\label{fig:quantum_token_process}
\end{figure}

Our paper aims to design a practical cloud-based semi-quantum money protocol. We design the protocol in the scenarios of token issuance, redemption, and circulation (shown in Figure \ref{fig:quantum_token_process}). Token issuance and redemption involve a token issuer, which we use a classical bank to represent. 

The bank and users are classical, meaning they can only perform classical computations and do not possess any quantum capabilities locally. This is a realistic assumption in the NISQ era where quantum computers are limited. To enable quantum functionalities, users and the bank can outsource quantum computations to the untrusted quantum cloud server. The cloud provider has a universal quantum computer but is not trusted. We assume it will follow the protocol honestly but try to learn information. Our protocol uses quantum homomorphic encryption to maintain privacy against the cloud.  By clearly defining the roles in our scenario, we can see both the bank and token holders are classical without needing a local quantum computer.

Algorithm \ref{alg: mint} shows how banks can mint a quantum token. This algorithm presents a protocol for minting a single quantum token \cite{shmueli2022public}. It involves three parties: a classical bank (Bank), a classical receiver (Rec), and a quantum third-party cloud server (Cloud). The goal of the algorithm is for the Bank to output a public key (pk), the Receiver to output a secret key (sk), and the Cloud to output a quantum state representing the quantum token (referred to as the \textbf{shifted quantum state}).

\begin{algorithm}[ht]
\footnotesize
\caption{Quantum Token Minting}\label{alg: mint}
\begin{algorithmic}[1] 

\REQUIRE A classical bank $Bank$, a classical receiver $Rec$, a quantum third-party cloud server $Cloud$, and a security paramater $\lambda$ 
\ENSURE In the end of the protocol, $Bank$ will output $pk := (O_{S_0+x}, O_{S_0+x+w}, O_{S^{\perp}+z})$, $Rec$ will output \{$R$, $X_R$, $Z_R$\}, and the $Cloud$ will output $(\ket{\$}_{pk})^{(R,0)} := (\ket{\psi}^{(x,z)})^{(R,0)}$ 

\STATE $Bank$ samples a random $\frac{\lambda}{2}$-dimensional subspace, described by matrix: $M_S \leftarrow \{0,1\}^{\frac{\lambda}{2}\times \lambda}$.

\STATE $Bank$ encrypts the matrix with random OTP key $p_x \leftarrow \{0,1\}^{\frac{\lambda}{2}\times \lambda}$, $M_S^{(p_x)} = QHE.OTP_{p_x}(M_S)$

\STATE $Bank$ encrypts the OTP key with $fhek \leftarrow QHE.Gen(1^{\lambda})$, $ct_{p_x} \leftarrow QHE.Enc_{fhek}(p_x)$. Then sends $(M_S^{(p_x)}, ct_{p_x})$ to $Rec$

\STATE $Rec$ samples shift key $R \leftarrow \{0,1\}^{\lambda}$, shifts per row in $M_S^{(p_x)}$, outputs $(M_{S+R}^{(p_x)} := (M_{S}^{(p_x)})_{ij} \oplus R_{j})$

\STATE $Rec$ sends $(M_{S+R}^{(p_x)}, ct_{p_x})$ to $Cloud$

\STATE $Cloud$ homomorphically evaluates $C: ( \ket{\psi+R}^{(x, z)}, ct_{(x, z)} ) \leftarrow QHE.Eval((M_{S+R}^{(p_x)}, ct_{p_x}), C)$ and sends back the access of $(\ket{\psi}^{(x,z)})^{(R,0)}$ and $ct_{x,z}$ to $Rec$

\STATE $Cloud$ appends an extra qubit $\ket{0}$ in the end of $(\ket{\psi}^{(x,z)})^{(R,0)}$

\STATE $Rec$ sends back $ct_{x,z}$ to $Bank$

\STATE $Bank$ decrypts $(x,z) = QHE.Dec_{fhek}(ct_{x,z})$.
\IF {$x\in S$}
$Bank$ terminates the interaction
\ELSE $Banks$ outputs the indistinguishability obfuscations $(O_{S_0+x}, O_{S_0+x+w}, O_{S^{\perp}+z})$ where $w$ is the first row of $M_S$, $S_0 \in \{0,1\}^{(\frac{\lambda}{2}-1 \times \lambda)}$ is the remain matrix of $M_S$ without $w$.
\ENDIF
\STATE $X_R$ = $(R;0)$
\STATE $Z_R$ = $0^{\lambda+1}$

\end{algorithmic}
\end{algorithm}

Digital token has high-security requirements. The shifted quantum state practice ensures that the cloud can not know the token information and operate the token in certain ways without the permission of token holders. We achieve this by using a two-phase quantum homomorphic encryption algorithm so that the cloud can conduct a quantum computation circuit along with encrypted data. Besides, replicating the same quantum token in a polynomial time is impossible because the generation process in the second step has a randomness that even the user can not produce the same quantum token with the information. The bank would only recognize the quantum token with $c_{x,z}$ received.

Algorithm \ref{Quantum Verification} shows how token holders can verify the authentication of quantum tokens. The receiver needs to know whether their token is valid and they can use the verification algorithm to conduct quantum computation with Oracle $pk$ that is outputted by the bank in the mining algorithm. The receiver can use the Oracle to delegate computational tasks to the cloud. If the result of the computation is 1, the token is valid. Otherwise, the token is not valid.

\begin{algorithm}[ht]
\footnotesize
\caption{Quantum Token Verification}\label{Quantum Verification}
\begin{algorithmic}[1] 

\REQUIRE A classical receiver $Rec$ holds \{$X_R$, $Z_R$\}, and a quantum third-party cloud server $Cloud$ stores $(\ket{\$}_{pk})^{(X_R, Z_R)} := (\ket{\psi}^{(x,z)})^{(X_R, Z_R)}$. 
\ENSURE In the end of the protocol, $Rec$ will output a new \{$X_R$, $Z_R$\}, and the $Cloud$ will output a new $(\ket{\$}_{pk})^{(X_R, Z_R)} := (\ket{\psi}^{(x,z)})^{(X_R, Z_R)}$ 

\STATE $Rec$ encrypts $X_R$, $Z_R$ with $reckey \leftarrow QHE.Gen(1^{\lambda})$, $ct_{(X_R,Z_R)} \leftarrow QHE.Enc_{reckey}((X_R,Z_R))$ and sends $ct_{(X_R,Z_R)}$, $(O_{S_0+x}, O_{S_0+x+w}, O_{S^{\perp}+z})$ to $Cloud$.
\STATE $Cloud$ homomorphically evaluates one of the oracle $O$ of $((O_{S_0+x} \vee O_{S_0+x+w}) \otimes I, (H^{\otimes \lambda}\otimes I)(O_{S^{\perp}+z})(H^{\otimes \lambda}\otimes I))$: $((\ket{\$}_{pk})^{(X'_R, Z'_R)}, ct_{(X'_R, Z'_R)}) \leftarrow QHE.Eval(((\ket{\$}_{pk})^{(X_R, Z_R)}, ct_{(X_R, Z_R)}), O)$ and sends the measurement result $\mu$ of the final qubit and $ct_{(X'_R, Z'_R)}$ to $Rec$
\STATE $Rec$ updates $X_R \leftarrow QHE.Dec_{reckey}(ct_{X'_R})$
\STATE $Rec$ updates $Z_R \leftarrow QHE.Dec_{reckey}(ct_{Z'_R})$
\STATE $Rec$ decrypts result $m\leftarrow QHE.Dec_{reckey}(\mu)$
\IF {$m=1$}
{Verification pass}
\ELSE {Verification fails}
\ENDIF
\STATE $Cloud$ and $Rec$ repeat to test the other oracle $O$
\IF {Both verifications passed}
{$Rec$ accepts the state}
\ELSE {$Rec$ rejects the state}
\ENDIF

\end{algorithmic}
\end{algorithm}

The quantum signing algorithm Algorithm \ref{Quantum Signing} allows a receiver to obtain a classical signature from a quantum state stored by a cloud provider while maintaining privacy. The receiver encrypts their secret keys and sends them to the cloud. The cloud uses homomorphic encryption to evaluate and sign the quantum state with a bit provided by the receiver. This enables the signing of quantum states without exposing the keys or the quantum state itself. The receiver decrypts the result to obtain the signature. This allows verifiable signing while keeping the state private.

The quantum transaction algorithm \ref{Quantum Transaction} enables two users with quantum tokens in separate clouds to transact. One user has a valued set of quantum tokens, as konw as cloud-base semi-quantum money, while the other has a set dummy tokens. By having the dummy state user send a unique bit string to the valued token holder, the real token holder can sign each quantum token and verify ownership to the other party without exposing the tokens. This allows value to be transferred from one user's tokens to the other, facilitated by the signature verification, without the need for the bank or direct access to the quantum states. Overall, the protocols allow for privacy-preserving transactions with quantum tokens in untrusted clouds via homomorphic encryption technology.

\begin{algorithm}[ht]
\footnotesize
\caption{Quantum Signing}\label{Quantum Signing}
\begin{algorithmic}[1] 

\REQUIRE A classical receiver $Rec$ holds \{$X_R$, $Z_R$\}, a bit $b \in \{0,1\}$, and a quantum third-party cloud server $Cloud$ stores $(\ket{\$}_{pk})^{(X_R, Z_R)} := (\ket{\psi}^{(x,z)})^{(X_R, Z_R)}$  
\ENSURE In the end of the protocol, $Rec$ will get a classical string $\sigma_b \in \{0,1\}^{\lambda}$. 

\STATE $Rec$ encrypts $X_R$, $Z_R$ with $reckey \leftarrow QHE.Gen(1^{\lambda})$, $ct_{(X_R,Z_R)} \leftarrow QHE.Enc_{reckey}((X_R,Z_R))$ and sends $ct_{(X_R,Z_R)}$, $(O_{S_0+x}, O_{S_0+x+w}, O_{S^{\perp}+z})$ to $Cloud$.
\STATE $Cloud$ homomorphically evaluates $((\ket{\$}_{pk})^{(X'_R, Z'_R)}, ct_{(X'_R, Z'_R)}) \leftarrow QHE.Eval(((\ket{\$}_{pk})^{(X_R, Z_R)}, ct_{(X_R, Z_R)}), O_{S_0+x+bw}\otimes I)$ and sends the measurement result $\mu$ of the final qubit to $Rec$ \label{QS.c}

\STATE $Rec$ updates $X_R \leftarrow QHE.Dec_{reckey}(ct_{X'_R})$ 
\STATE $Rec$ updates $Z_R \leftarrow QHE.Dec_{reckey}(ct_{Z'_R})$
\STATE $Rec$ decrypts result $m\leftarrow QHE.Dec_{reckey}(\mu)$
\IF {$m=1$}
{$Cloud$ measures the register $\ket{\$}_{pk})^{(X_R, Z_R)}$ to get measurement $(\sigma_b)' \leftarrow \sigma_b \oplus X_R$, sends $(\sigma_b)'$ to $Rec$ and terminate.}
\ELSE {Perform Algo. \ref{Quantum Verification} of the oracle $(H^{\otimes \lambda}\otimes I)(O_{S^{\perp}+z})(H^{\otimes \lambda}\otimes I)$ and return to step \ref{QS.c}}
\ENDIF
\STATE $Rec$ calculate $\sigma_b \leftarrow (\sigma_b)' \oplus X_R $

\end{algorithmic}
\end{algorithm}

\begin{algorithm}[ht]
\footnotesize
\caption{Quantum Transaction}\label{Quantum Transaction}
\begin{algorithmic}[1] 

\REQUIRE A classical sender $A$ who can access a $\lambda$-length sequence of quantum token $Token_A := (\ket{\$_1}_{pk_1})^{(X_{R_1},Z_{R_1})}, \ket{\$_2}_{pk_2})^{(X_{R_2},Z_{R_2})},..., \ket{\$_\lambda}_{pk_\lambda})^{(X_{R_\lambda},Z_{R_\lambda})})$ stored in $Cloud_A$ and $x_A, s_A, \sigma_A$, where $x_A$ is the value of $Token_A$, $s_A \in \{0,1\}^\lambda$ is a unique bit string and $\sigma_A$ is the classical signature of $((pk_1, pk_2,...,pk_\lambda)_A, x_A, s_A)$. A classical receiver $B$ can access a dummy $\lambda$-length sequence of quantum token $Token_B$  stored in $Cloud_B$, $x_B = 0$. 
\ENSURE In the end of the protocol, the value $x_A$ of $Token_A$ will be added to the value $x_B$ of $Token_B$. 

\STATE $B$ sends $A$ the series number $s_B \in \{0,1\}^{\lambda}$
\STATE $A$ signs each token in $Token_A$ with the bit from $s_B$ respectively: $\sigma_{B \leftarrow A} := (\sigma_{s_{B_1}}, \sigma_{s_{B_2}},..., \sigma_{s_{B_\lambda}})_{Token_A}$ and sends the signature $\sigma_{B \leftarrow A}$ to $B$

\end{algorithmic}
\end{algorithm}

To assign actual value to QT. The user will pay a certain price to the bank to obtain a string of QT sequence of length $\lambda$ issued by the bank to the user, a unique random sequence number $s$, the value of this string of QT $x$, and a classic signature to prove the integrity of this string of QT $\sigma$ (that is to say, $\sigma$ can guarantee that this string of QT can be used normally if and only if its serial number $s$ and value $x$ are all consistent with the state when they were issued).

\subsection{Mint a Quantum Token Unit}
In this section, we give out the mint circuit of the basic unit of the quantum token along with a 4-qubit demonstration for clear.
\subsubsection{Encrypted Space Expansion Circuit}
\begin{figure*}[bpt]
    \centering
    \includegraphics[width=1\textwidth]{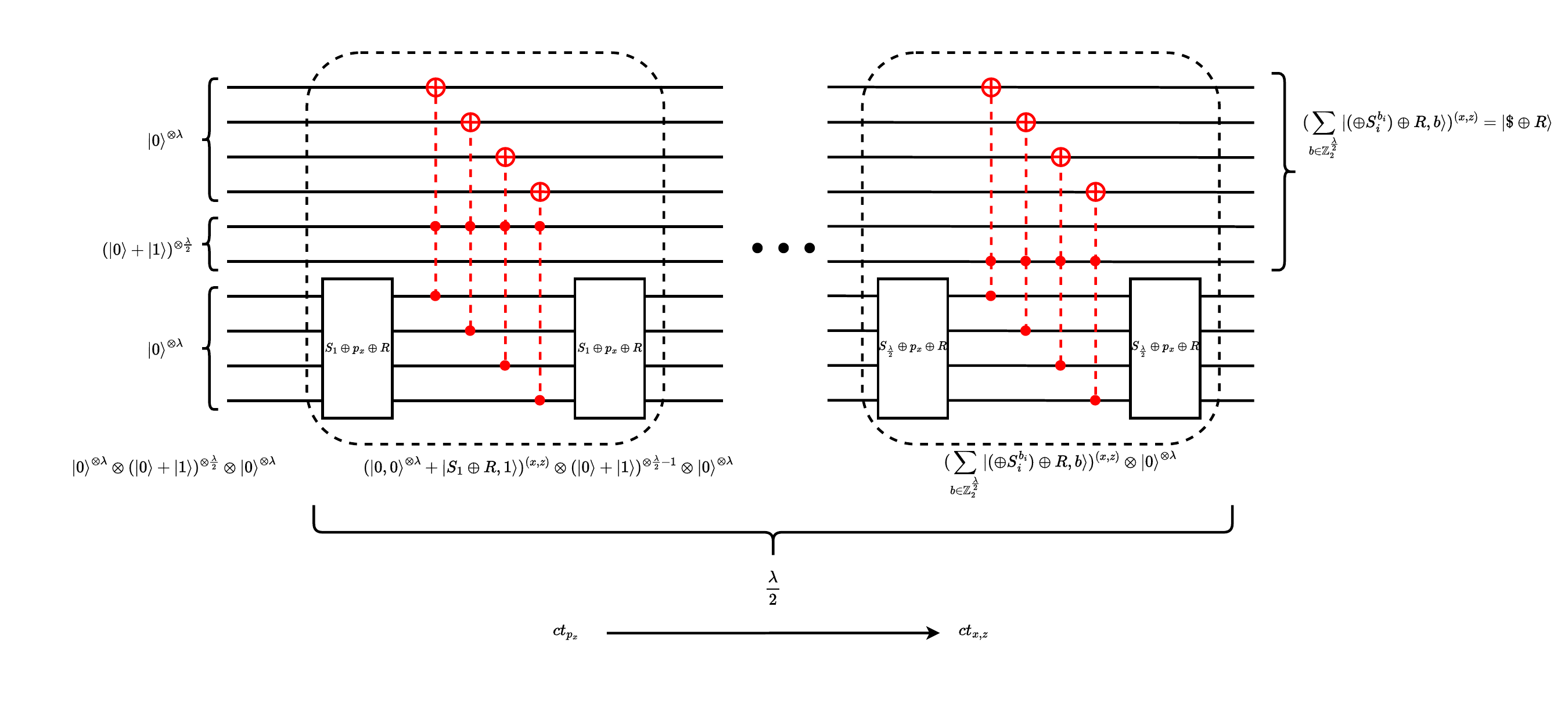}
    \caption{Mint Circuit of a Quantum Token Unit}
    \label{mint}
\end{figure*}

In Algo. \ref{alg: mint}, the cloud $C$ receives $(M_{S+R}^{(r)}, ct_r)$ and the mint circuit $C_m$ from the classical receiver $R$. The mint circuit $C_m$ (Figure \ref{mint}) is constructed by $\frac{\lambda}{2}$ blocks. Each block will add one new basis into the space (described by a quantum state) and in the end of the circuit, it will output the OTP encrypted space row span of $M_{S+R}$, also can be written as $(\sum_{b \in \mathbb{Z}_2^\frac{\lambda}{2}}\ket{(\oplus S_i^{b_i})\oplus R,b})^{(x,z)}=\ket{\$\oplus R}$, where $S_i$ is the $i$-th row of $M_S$. In addtion, the classical ciphertext $ct_{(x,z)}$ transforms into $ct_{(x',z')}$.

\subsubsection{4-Qubit Demonstration}
\begin{figure*}[bpt]
    \centering
    \includegraphics[width=1\textwidth]{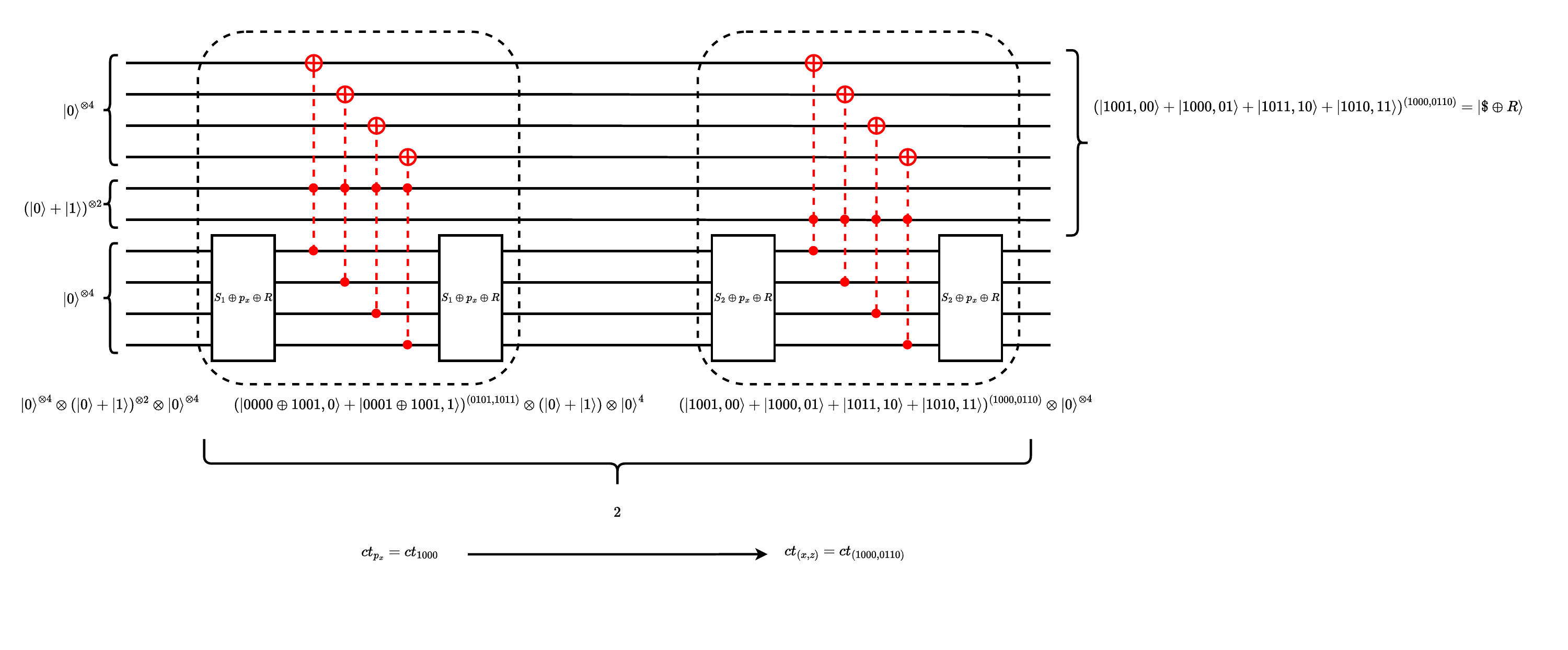}
    \caption{Demonstration of a 4-Qubit Quantum Token Unit}
    \label{mint_demo}
\end{figure*}

In this section, we suggest a 4-qubit demonstration for clear description (Figure \ref{mint_demo}). We set the parameters as:
\begin{enumerate}
    \item $\lambda=4$
    \item $M_S = \begin{bmatrix}
0 & 0 & 0 & 1\\
0 & 0 & 1 & 0
\end{bmatrix}$
    \item $p_x = 1000$
    \item $R=1001$
\end{enumerate}
Let's review Algo. \ref{alg: mint}. The Bank sends Receiver $M_S^{p_x} = \begin{bmatrix}
1 & 0 & 0 & 1\\
1 & 0 & 1 & 0
\end{bmatrix}$. And the Receiver applies $R$ on $M_S^{p_x}$ to earn $M_{S+R}^{p_x} = \begin{bmatrix}
0 & 0 & 0 & 0\\
0 & 0 & 1 & 1
\end{bmatrix}$ and send it to the cloud $C$. After homomorphically run the mint circuit $(\ket{\$ + R}, ct_{(1000, 0110)} \leftarrow QHE.Eval(C_m, \begin{bmatrix}
0 & 0 & 0 & 0\\
0 & 0 & 1 & 1
\end{bmatrix}, ct_{1000}))$, the cloud $C$ will hold the $R$-encrypted quantum token $\ket{\$ + R}= (\ket{0000\oplus 1001,00}+\ket{0001\oplus 1001,01}+\ket{0010\oplus 1001,10}+\ket{0011\oplus 1001,11})^{(1000,0110)}$ and $ct_{(1000, 0110)}$.

\subsubsection{Quantum Verification and Quantum Signing}
\begin{figure}
    \centering
    \includegraphics[width = 0.5 \textwidth]{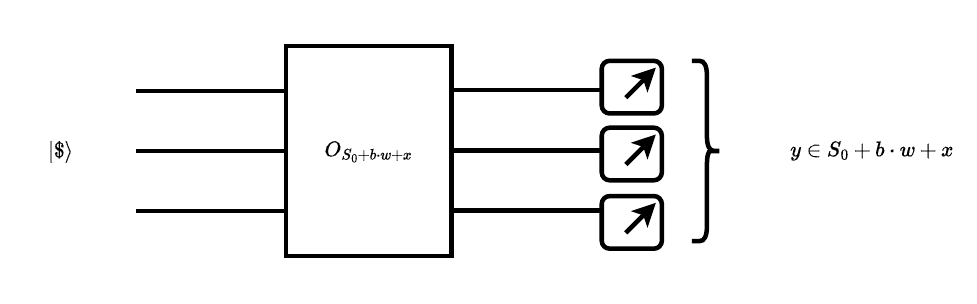}
    \caption{Quantum Signing}
    \label{O}
\end{figure}

After mint process, the bank will public three oracles $(O_{S_0+x}, O_{S_0+x+w}, O_{S^{\perp}+z})$ to advise everyone how to verify the quantum token $\ket{\$} = \sum_{s \in S}\ket{s}^{(x',z')}$ by the measurement results through the mint process \cite{shmueli2022semi}. In other words, since the banks holds the trapdoor $t_r$, it can reveal the exact value of $(x',z')$. We define $S_0$ is a $(\frac{\lambda}{2}-1)$-dim subspace of $S$ and $S_0 + w$ is the coset of $S_0$. In the quantum signing protocol (Figure. \ref{O}, the sender provide an arbitrary bit $b \in \{0,1\}$ and the prover who holds the quantum token $\ket{\$}$ can sign the token with $b$. Finally, the prover outputs $\sigma_b = y\in S_0 +b\cdot w + x$. In Algo. \ref{Quantum Signing}, the measurement is homomorphically executed by the cloud $C$ with the assistant of $R$.

In this section, we delve into the analysis of the security aspects of our delegation procedure in Cloud Quantum Computation. Our foundational assumption is that the cloud, denoted as $C$, acts as a semi-honest third party that possesses quantum capabilities.

\section{Discussion}

\subsection{Claim on Quantum State Acquisition}

Under the assumption of the cloud $C$ is fault  tolerated. We assert that after our mint protocol, the cloud $C$ successfully acquires a specific set of quantum states. These states are particularly significant because they exhibit a negligible difference, in terms of trace distance, from a defined quantum state denoted as $\ket{\Psi + R}^{(x,z)}$.

The bedrock of this claim lies in the statistical security inherent in Quantum Fully Homomorphic Encryption (QFHE). This security assurance leads to a high probability - essentially certainty - that the cloud's register will indeed contain quantum states. These states, as previously mentioned, will have a negligible trace distance from our specified state $\ket{\Psi + R}^{(x,z)}$.

In the quantum token verification and signing part, receiver $Rec$ sends $ct_{(X_R,Z_R)}$, $(O_{S_0+x}, O_{S_0+x+w}, O_{S^{\perp}+z})$ to $C$. $C$ homomorphically evaluates one of the oracle $O$ of $((O_{S_0+x} \vee O_{S_0+x+w}) \otimes I, (H^{\otimes \lambda}\otimes I)(O_{S^{\perp}+z})(H^{\otimes \lambda}\otimes I))$: $((\ket{\$}_{pk})^{(X'_R, Z'_R)}, ct_{(X'_R, Z'_R)}) \leftarrow QHE.Eval(((\ket{\$}_{pk})^{(X_R, Z_R)}, ct_{(X_R, Z_R)}), O)$ and sends the measurement result $\mu$ of the final qubit and $ct_{(X'_R, Z'_R)}$ to $R$. If $C$ performs QFHE correctly, the output state will have a negligible trace distance from $\ket{\$}_{pk})^{(X_R, Z_R)}$.

\subsection{Claim on the Secrecy of Shift $R$}

A pivotal aspect of our protocol is the confidentiality of the shift, denoted as $R \in \{0,1\}^{\lambda}$. Our claim here is that this shift remains a secret throughout the entire protocol.

We consider a hypothetical scenario where the cloud $C$ can manipulate another quantum token, say $\ket{\Psi'}$, in such a way that it passes the quantum verification process (QV) in polynomial time. In such a case, $C$ would be able to decrypt a randomly selected secret $S$. This possibility, however, leads to a contradiction in terms of the security of one-time-pad, thereby underscoring our claim that the shift $R$ remains confidential throughout the process.
\subsection{Estimation of Quantum Resources}
We set the key size $\lambda$ is 256 bits to ensure the security of the quantum token. The plain space expansion circuit of a $\frac{\lambda}{2}$-dim space requires $2.5\lambda$ qubits. The Toffoli gate of QFHE in the circuit needs at least $\lambda$ qubits. Therefore our protocol requires quantum computation resources with $~900$ qubits.


\bibliography{refs}
\bibliographystyle{unsrt}

\end{document}